# Observation of even denominator fractional quantum Hall effect in suspended bilayer graphene

*Dong-Keun Ki[1], Vladimir I. Fal'ko[2], Dmitry A. Abanin[3,4], and Alberto F. Morpurgo[1*]*

[1]Départment de Physique de la Matiére Condensée (DPMC) and Group of Applied Physics (GAP), University of Geneva, 24 Quai Ernest-Ansermet, CH1211 Genéve 4, Switzerland.

[2]Department of Physics, Lancaster University, Lancaster, LA1 4YB, UK.

[3]Perimeter Institute for Theoretical Physics, Waterloo, Ontario N2L 2Y5, Canada

[4]Institute for Quantum Computing, Waterloo, Ontario N2L 3G1, Canada

We investigate low-temperature magneto-transport in recently developed, high-quality multi-terminal suspended bilayer graphene devices, enabling the independent measurement of the longitudinal and transverse resistance. We observe clear signatures of the fractional quantum Hall effect, with different states that are either fully developed, and exhibit a clear plateau in the transverse resistance with a concomitant dip in longitudinal resistance, or incipient, and exhibit only a longitudinal resistance minimum. All observed states scale as a function of filling factor $\nu$, as expected. An unprecedented even-denominator fractional state is observed at $\nu = -1/2$ on the hole side, exhibiting a clear plateau in $R_{xy}$ quantized at the expected value of $2h/e^2$ with a precision of ~0.5%. Many of our observations, together with a recent electronic compressibility measurement performed in graphene bilayers on hexagonal boron-nitride (hBN) substrates, are consistent with a recent theory that accounts for the effect of the degeneracy between the *N=0* and *N=1* Landau levels in the fractional quantum Hall effect, and predicts the occurrence of a Moore-Read type $\nu = -1/2$ state. Owing to the experimental flexibility of bilayer graphene –which has a gate-dependent band structure, can be easily accessed by scanning probes, and can be contacted with materials such as superconductors–, our findings offer new possibilities to explore the microscopic nature of even-denominator fractional quantum Hall effect.

In the presence of a strong magnetic field (*B*), the behavior of ultra-clean two-dimensional electron systems (2DESs) is determined by electronic interactions and quantum mechanical correlations, which results in a rich variety of so-called fractional quantum Hall states.[1,2] In transport measurements, these states manifest themselves through plateaus in the transverse resistance $R_{xy} = 1/\nu \times h/e^2$ at fractional values of $\nu$ ($\nu \equiv nh/eB$ is the filling factor, i.e. the ratio between the carrier density *n* and the Landau level degeneracy; *h* is Planck's constant), occurring concomitantly with dips in the longitudinal resistance $R_{xx}$.[1,2] In the vast majority of cases, fractional quantum Hall effect (FQHE) occurs when $\nu$ is an odd-denominator fraction,[1-7] as described first by Laughlin,[8] and later by Jain, who introduced composite fermion theory.[9] Only in very rare cases, even-denominator FQHE is observed.[2,10-12] The nature of these even-denominator states is poorly understood, and is attracting interest because some of them are expected to support Majorana zero-modes with non-Abelian statistics.[13-22] Here, we demonstrate a clear occurrence of an even-denominator FQHE at $\nu = -1/2$ on the hole side in bilayer graphene, and show that our experimental observations are consistent with a recently proposed theory[23] that predicts the $\nu = -1/2$ state to be of the Moore-Read type.

The nature of fractional quantum Hall states depends strongly on how electronic interactions break symmetries and degeneracies in a 2DES.[2] The role of interaction-induced symmetry breaking in FQHE has been studied in GaAs-based heterostructures,[2] and in monolayer graphene,[24, 25] where the FQHE was observed at an unusual sequence of odd-denominator filling fractions, possibly determined by an approximate SU(4) symmetry associated with the spin and valley degrees of freedom.[26-28] Comparing FQHE in mono- and bi-layer graphene –consisting of two Bernal-stacked monolayers– is particularly revealing,[23, 29-33] since in the absence of interactions the low-energy structure of Landau levels is very different in the two cases (despite the spin and valley degeneracy being identical).[34, 35] Specifically, whereas only one $N=0$ Landau level is present at zero energy in monolayers, in bilayers $N=0$ and $N=1$ Landau level both have vanishing energy, leading to a unique additional degeneracy in the system[34-42] (conventional GaAs-based systems do not have this extra degeneracy). This difference is crucial as it leads to unexplored regimes of electronic interactions in 2DESs where mixing of two $N=0$ and $N=1$ Landau levels is important,[23, 29-33] and for which a partially filled $N=1$ (rather than $N=0$) Landau level occurs already at low carrier density.

So far, a clear experimental observation of FQHE in transport measurements has been missing in bilayer graphene (the occurrence of a state at $\nu = 1/3$ was claimed in suspended bilayer graphene, but the reported experimental evidence was weak, owing to the lack of proper conductance quantization).[43] This is because, in contrast to the monolayer case, in bilayers the characteristic scale of the energy gap protecting the stability of the FQHE is an order of magnitude smaller, making the FQHE much more susceptible to disorder. Only very recently, signatures of the FQHE in bilayer graphene were observed in devices based on hexagonal boron nitride (hBN) substrates, through local electronic compressibility measurements relying on a scanning probe method.[44] Indeed, by probing only local properties, this technique enables a sufficiently clean area to be found where disorder effects are small, even in cases when the overall device quality is insufficient to observe the FQHE in transport measurements.

A key to obtain the results reported here is the ability to identify quantum Hall features unambiguously in high-quality suspended graphene devices, by separately measuring plateaus in the transverse resistance $R_{xy} = 1/\nu \times h/e^2$, and accompanying dips in the longitudinal resistance $R_{xx}$. The majority of transport experiments on suspended graphene devices performed in the past to probe the quantum Hall effect were carried out in a two-terminal configuration,[6, 7, 38, 43, 45] and did not allow the independent determination of $R_{xx}$ and $R_{xy}$ (see Ref. 46 for an exception). However, recently developed multi-probe suspended devices[47] offer new possibilities thanks to their unprecedented uniformity and very high mobility values ($\mu \sim 10^6$ cm$^2$/V/s; see Supporting Information and Ref. 47 for details of the fabrication and annealing process, and for extensive discussion of the device characterization). Here we discuss measurements performed on two of these devices, exhibiting virtually identical features associated to different fractional quantum Hall states (the data shown in the main text have been taken on one of the devices; data from the second device are shown in the Supporting Information).

Figure 1a shows the magneto-resistances, $R_{xx}$ ($\equiv R_{12,43}$) and $R_{xy}$ ($\equiv R_{13,42}$), measured at a back-gate voltage $V_{BG} = -27$ V (at $n = \alpha \times (V_{BG} - V_{CNP}) \approx -1.64 \times 10^{11}$ /cm$^2$, with $\alpha \approx 5.29 \times 10^9$ /cm$^2$/V and with a charge neutrality at $V_{CNP} \approx 4$ V obtained by scaling the quantum Hall features at integer filling factors below 1 T; Supporting Information). At rather low field, the integer quantum Hall states at $\nu = -1$, -2, and -3 give origin to very wide plateaus in $R_{xy}$ quantized exactly at $1/\nu \times h/e^2$ with fully vanishing $R_{xx}$, as expected, implying the complete breaking of spin and valley degeneracy of the zero-energy Landau levels.[36-42] This high quality of integer QHE which survives down to 30 mT (see Supporting Information for extensive sets of low-$B$ QHE data) establishes the reliability of our measurements in identifying fractional quantum Hall features that are also visible in the measurements (the data also demonstrate a very low level of inhomogeneous doping, a few times $10^9$ /cm$^2$, excluding possible artifacts due to the presence of regions in the device with different doping in our measurements of the integer and fractional QHE; see Supporting Information for more details). The most apparent of these features are a plateau in $R_{xy} = 1/\nu \times h/e^2$ with $\nu = -1/2$ ($\approx 51.5$ k$\Omega$), appearing at B $\approx 13.5$ T (Figure 1b), and a plateau at $R_{xy} = 1/\nu \times h/e^2$ with $\nu = -4/3$ ($\approx 19.2$ k$\Omega$) at B $\approx 5$ T (Figure 1c), both accompanied by a minimum in $R_{xx}$. Additional features are present in the data, and in order to identify those associated to FQHE we investigate the full dependence of $R_{xx}$ on $V_{BG}$ and $B$ (Figure 2a).[27, 44]

When $V_{BG}$ is changed, the fractional features in Figure 1a appear at different values of $B$ (see Figure 2a), and exhibit a linear dispersion as expected for the quantum Hall states that occur at fixed filling factor $\nu$. We therefore plot the data as a function of $\nu$ and $B$, so that the features originating from the QHE become "vertical", and can be readily identified.[27, 44] This is done in Figure 2b on the hole side ($V_{BG} < V_{CNP}$) where, next to the minima at $\nu = -1/2$ (see Figure 2c for a zoom in) and -4/3, a pronounced minimum in $R_{xx}$ at $\nu = -8/5$ is visible, together with an additional one at $\nu = -2/3$ (partially eclipsed by disorder; Supporting Information). A pronounced minimum in $R_{xx}$ at $\nu = -5/2$ is also clearly apparent in the data (see Figure 2d for a zoom in), as well as a less pronounced one at $\nu = 2/3$ on the electron side (Supporting Information). As they occur at a fixed fractional filling factor, all these features (observed in both devices investigated) are indicative of existing fractional quantum Hall states having an energy gap that is not sufficiently large to enable the observation of a fully developed plateau in $R_{xy}$. For the most robust states at $\nu = -1/2$ and $-4/3$ –for which a well-developed plateau in $R_{xy}$ is observed– we analyze the data, by plotting measurements at different values of $B$ and $V_{BG}$ as a function of $\nu$ on a same graph, to check whether all curves fall on top of each other. The unambiguous scaling of the plateaus in $R_{xy} = 1/\nu \times h/e^2$ (with values of $\nu$ that agree with $\nu = -1/2$ and $-4/3$ within 0.2% and 0.7% respectively, as shown in Figures 3b and 3e) and of the dips in $R_{xx}$ is apparent (Figure 3).[27, 44] Finally, the multi-terminal geometry of our devices allows us to estimate the characteristic energy scale of the fractional quantum Hall states (determined, in the ideal case, by the corresponding energy gap) by looking at the temperature dependence of minimal values of $R_{xx}$, which is expected to follow thermally activated law, $R_{min} = \exp(-\Delta_\nu/2k_BT)$ (Supporting Information).[26, 46] For the state at $\nu = -1/2$ (-4/3), the $R_{xx}$ curves measured as a function of $\nu$ at different temperatures are shown in Figure 4a (4b), and Figure 4c summarizes the magnetic field dependence of the energies, $\Delta_{-1/2}$ (blue circle) and $\Delta_{-4/3}$ (red squares) which increase smoothly with increasing $B$.

To summarize the experimental results, we observe the occurrence of fractional quantum Hall states –either fully developed or incipient– at a sequence of fractions $\nu = -1/2, -2/3, -4/3, -8/5$, and $-5/2$ for holes and at $\nu = 2/3$ for electrons, which has been found to be the same in the two samples that we have investigated in detail. This sequence, therefore, can be taken as a fingerprint of the FQHE in bilayer graphene, i.e. of the states in this system that are most robust. Among them, the states at $\nu = -1/2$ and $-4/3$ are fully developed, i.e. they exhibit a clear plateau in $R_{xy}$ (with the $\nu = -4/3$ state having the largest energy gap; see Figure 4c). This behavior differs from what is found in other 2DESs, including monolayer graphene, where only odd-denominator states have been observed with the strongest state at $|\nu| = 1/3$.[6, 7, 26, 27, 46] It is also clearly distinct from that of double quantum-well system consisting of two weakly coupled 2DESs by Coulomb repulsion, where 1/2

FQHE was found:[11, 12] indeed, the constituent monolayers of our bilayer graphene are strongly tunnel coupled, with an inter-layer hopping integral $t^* \sim 400$ meV,[34, 35] and behave as an individual 2DES (see Supporting Information). Such unique sequence of fractions, therefore, can be used to gain insight about the nature of fractional quantum Hall states, particularly of the one at $\nu = -1/2$, through comparison with theory.[23]

Recently, the problem has indeed been addressed theoretically,[23, 29-33] and the analysis of Ref. 23 is particularly relevant. This theory takes into account the fact that the zero-energy Landau level in bilayer graphene has a unique eight-fold degeneracy which is lifted by the exchange interaction, so that spin and valley degeneracies are broken first, followed by the splitting of the orbital states, $N=0$ and $N=1$ (see the inset of Figure 4c).[34-42] As a consequence, in bilayer graphene, the Coulomb interaction can scatter electrons with the same spin/valley indices between these two levels,[23, 29-33] which is neither possible in conventional GaAs-based systems[2] nor in monolayer graphene (due to the large cyclotron gap).[6, 7, 24-27, 46] Numerical calculations based on this scenario show that the states between filling factors $2k < \nu < 2k+1$ ($k = -2, -1, 0,$ and $1$) correspond to a partially filled $N=0$ Landau level, while those between $2k+1 < \nu < 2k+2$ correspond to a partially filled $N=1$ Landau level (with different spin/valley quantum numbers for the different values of $k$).[23] This scheme automatically predicts an approximate $\nu \rightarrow \nu+2$ symmetry and accounts for the presence of electron-hole asymmetry: starting from charge neutrality, electrons are removed from a $N=1$ Landau level while they are added to a $N=0$ Landau level (see inset of Figure 4c). Theory also predicts that the most robust state in the filling factor interval $2k < \nu < 2k+1$ (i.e., when a $N=0$ orbital is being filled) occurs at $\nu = 2k+2/3$, with weaker states expected at $\nu = 2k+2/5$ and $2k+1/3$. Finally, and particularly interestingly, at $2k+1 < \nu < 2k+2$ –i.e., when a $N=1$ Landau level is being filled– incompressible fractional quantum Hall states are expected at half-integer $\nu = 2k+3/2$, whose ground states are identified as non-Abelian Moore-Read states[13] (i.e., the calculated ground state has a large overlap with the Moore-Read state, much larger than, for instance, with the 331 state).

We interpret our results within the context of these theoretical predictions, whose validity is supported by the recent local probe electron compressibility measurements performed on bilayers on hBN substrate that we referred to above,[44] which directly shows the occurrence of the $\nu \rightarrow \nu+2$ symmetry of FQHE. We find an overall good agreement with theoretical predictions, starting from the value of filling factor and of polarity of the most robust states: we observe the two more robust states at $\nu = -4/3$ (= $2k+2/3$ for $k = -1$) and at $\nu = -1/2$ (= $2k+3/2$ for $k = -1$) on the hole side, just as predicted by the theory. Additionally, signatures of the states at $\nu = -5/2$ (= $-1/2-2$) and $\nu = 2/3$ (= $-4/3+2$) are observed, as expected for the $\nu \rightarrow \nu+2$ symmetry, and the weaker state at $\nu = -8/5$, (= $2k+2/5$ for $k = -1$; one of the "second-most-robust" states according to theory) is detected experimentally as well. There are, however, also differences between theory and experiments, the most important of which is the near absence of fractional features on the electron side, even though the features from integer QHE appear symmetrically on both sides (Supporting Information). We attribute this fact to the presence of remnant disorder (note that the gaps protecting the fractional states are small, and charge inhomogeneity that would be totally inconsequential for integer quantum Hall states can suppress the FQHE), similarly to the case of previous transport experiments on suspended graphene monolayers, in which fractional features in transport measurements have also been reported for only one polarity of charge carriers.[6, 7] Indeed, in graphene systems, disorder close to charge neutrality has been often found not to be electron-hole symmetric, with the asymmetry becoming quite pronounced at high magnetic field[7] (another possibility to consider is that doping inhomogeneity in the graphene leads, which is also somewhat larger on the electron side, may play a role in preventing the observation of FQHE on the electron side; although, quantitatively, the inhomogeneity seems to be significantly too small to cause large effects, it is worth keeping in mind its presence in the devices; see Supporting Information)

Clearly the most notable experimental result in our data is the occurrence of the $\nu = -1/2$ state that is theoretically predicted to be of the Moore-Read type,[23] which had never been detected in previous experiments. The reason why such a state could not be detected in the electron compressibility experiments[44] is probably related to its energy scale. In our experiments on suspended graphene the state appears at $T \sim 1$ K and the quantized plateau in $R_{xy}$ is fully developed already between 250 and 500 mK. However, the dielectric constant of hBN substrates ($\varepsilon = 4$) screens electron-electron interaction and suppresses the gap protecting the $\nu = -1/2$ state by $(\varepsilon + 1)/2 = 2.5$. This suppression may explain why the electron compressibility experiments, that were performed at a temperature of $T = 450$ mK,[44] nearly twice our base temperature, the $\nu = -1/2$ fractional quantum Hall state has not been detected. Irrespective of these considerations, we emphasize that as compared to most experiments on the 5/2 state in GaAs-based heterostructures –which typically require $T$ to be much smaller than 100 mK– the temperature at which the $\nu = -1/2$ state in bilayer graphene is experimentally visible in our experiments is one order of magnitude higher.

We believe that our results have a number of important implications. One is the possibility to investigate the universality of even-denominator fractional quantum Hall states, and in particular the physics of a half-filled $N=1$ Landau level, through a detailed comparison of the 5/2 state in GaAs and the 1/2 state in bilayer graphene. Graphene bilayers are attractive in this regard because their microscopic properties can be tuned by a perpendicular electric field to unbalance the two layers or to control symmetries,[23, 30-33, 48, 49] and because (as seen in this experiment) accessing the $\nu = -1/2$ state is possible at relatively high temperature (~ 0.5 K). Another aspect is the possibility to access the 2DES in bilayer graphene readily with different kinds of contacts, including superconducting[50] and ferromagnetic ones,[51] and with scanning probes,[27, 28, 44] all of which can be used to investigate the nature of this even-denominator state in new ways. Our work therefore opens the path to unprecedented studies of even-denominator fractional quantum Hall effect, and raises the exciting possibility to use bilayer graphene as a suitable experimental platform for studying the physics of non-Abelian excitations.

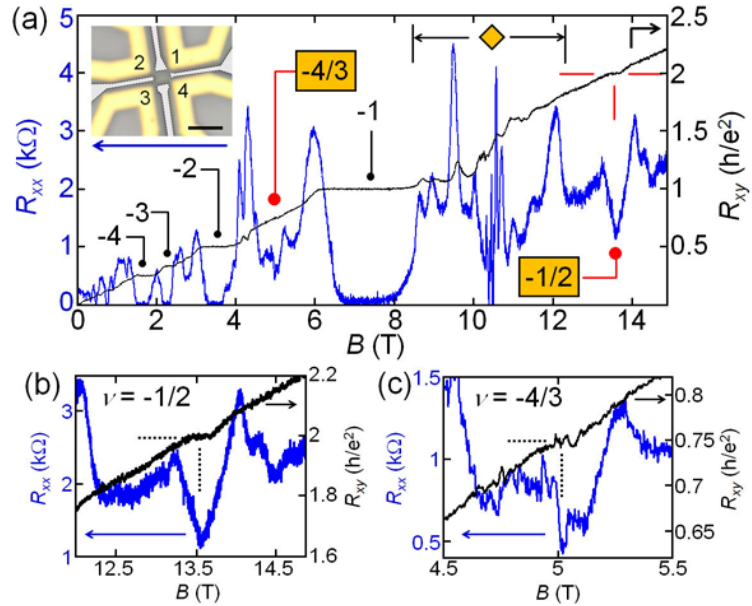

**Figure 1.** Observation of FQHE at $T = 0.25$ K. (a) Magneto-resistance $R_{xx}$ ($R_{12,43}$, blue curve) and $R_{xy}$ ($R_{13,42}$, black curve) at $V_{BG} = -27$ V, displaying rich features from integer and fractional QHE, pointed to by red and black lines. Here, $R_{\alpha\beta,\gamma\delta}$ = (voltage measured between electrodes $\gamma$ and $\delta$)/(current flowing from $\alpha$ to $\beta$). The features in the range marked by the yellow diamond are affected by remnant disorder (see also Figure 2a). The inset shows an image of the actual device covered with a resist mask. Metallic electrodes are labeled by numbers; graphene is etched away in the grey area not covered by the resist. The scale bar is 3 µm long. (b, c) Zoom-in on the data

shown in panel a, to better illustrate the features associated with the -1/2 and -4/3 fractional quantum Hall states, respectively.

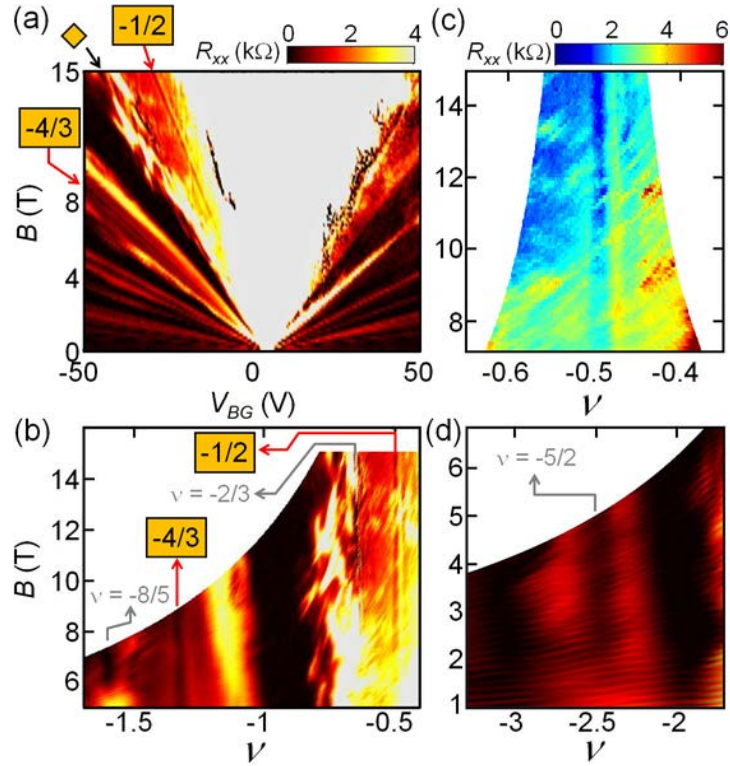

**Figure 2.** Identification of fractional quantum Hall features in Landau fan-diagram at $T = 0.25$ K. (a) Color plot of $R_{xx}$ as a function of $V_{BG}$ and $B$. The bright region around $V_{BG} = V_{CNP}$ is a magnetic field induced insulating state that is commonly seen in high-quality graphene mono- and bi-layer. The linearly dispersed dark stripes associated to quantum Hall states –local minima in $R_{xx}$– become vertical (b) at $\nu =$ -1/2, -2/3, -4/3, and -8/5 when plotted as a function of $\nu$. (c, d) Zoom-in on the data around $\nu =$ -1/2 and -5/2, respectively, showing a clear local minimum in $R_{xx}$ as a vertical strip at the center.

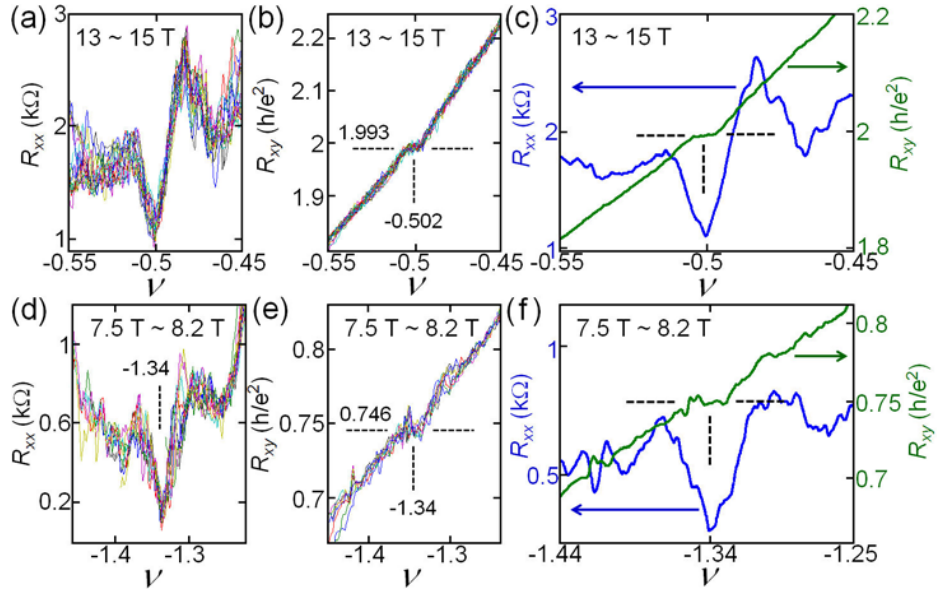

**Figure 3.** Scaling of magnetotransport in fractional quantum Hall regime at $T = 0.25$ K. (a-c) The measurements around $\nu = -1/2$ for $B$ in the range 13-15 T. The dips in $R_{xx}$ (a) and the plateaus in $R_{xy}$ (b) clearly collapse on top of each other when plotted as a function of $\nu$. The observed good quality of scaling enables averaging over the different traces to decrease the experimental noise (c). (d-f) Data measured around $\nu = -4/3$ for $B$ in the range 7.5-8.2 T plotted in the same way as in (a-c).

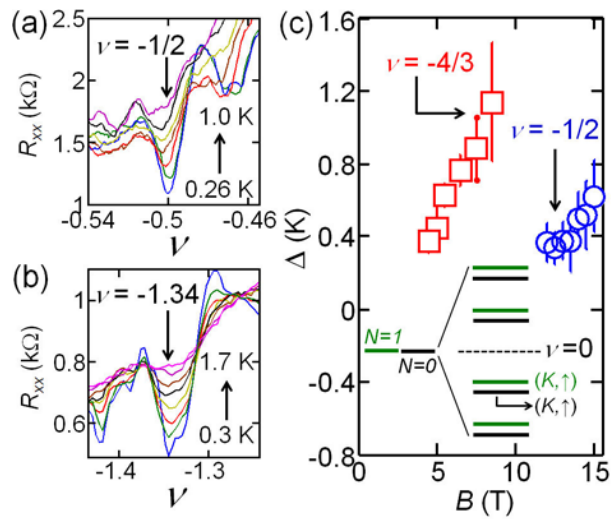

**Figure 4.** Activation energies of the -1/2 and -4/3 fractional quantum Hall states. (a, b) $R_{xx}(\nu)$ measured around $\nu$ = -1/2 (at $B$ = 15 T) and -4/3 (at $B$ = 8.5 T), respectively, for increasing temperature $T$ (from bottom to top). (c) Monotonic increase of the measured gap energies, $\Delta_{-1/2}$ (blue circle) and $\Delta_{-4/3}$ (red square), with increasing magnetic field $B$. The schematic drawing in the inset shows the complete splitting of the zero-energy Landau level of bilayer graphene that provides the starting point for the theoretical analysis discussed in the main text.

## ASSOCIATED CONTENT

**Supporting Information.** In addition to the fabrication details and extensive discussion of the device characterization, the results from different device and the evidence of weaker fractional quantum Hall states are also provided. This material is available free of charge via the Internet at http://pubs.acs.org.

## AUTHOR INFORMATION

**Corresponding Author**

E-mail: *Alberto.Morpurgo@unige.ch

**Notes**

The authors declare no competing financial interest.

## ACKNOWLEDGMENT
We acknowledge A. Ferreira for technical support, Z. Papić for theoretical discussion, and A. Grushina for her assistance with the measurements. Financial support from the SNF, NCCR MaNEP, NCCR QSIT, and EU Graphene Flagship are gratefully acknowledged.

## REFERENCES

(1) Tsui, D. C.; Stormer, H. L.; Gossard, A. C. *Phys. Rev. Lett.* **1982,** 48, (22), 1559-1562.

(2) Yoshioka, D., *The quantum Hall effect*. Springer: Berlin ; New York, 2002; p xii, 207 p.

(3) Nelson, S. F.; Ismail, K.; Nocera, J. J.; Fang, F. F.; Mendez, E. E.; Chu, J. O.; Meyerson, B. S. *Appl. Phys. Lett.* **1992,** 61, (1), 64-66.

(4) Lai, K.; Pan, W.; Tsui, D. C.; Lyon, S.; Mühlberger, M.; Schäffler, F. *Phys. Rev. Lett.* **2004,** 93, (15), 156805.

(5) Tsukazaki, A.; Akasaka, S.; Nakahara, K.; Ohno, Y.; Ohno, H.; Maryenko, D.; Ohtomo, A.; Kawasaki, M. *Nat. Mater.* **2010,** 9, (11), 889-893.

(6) Du, X.; Skachko, I.; Duerr, F.; Luican, A.; Andrei, E. Y. *Nature* **2009,** 462, (7270), 192-195.

(7) Bolotin, K. I.; Ghahari, F.; Shulman, M. D.; Stormer, H. L.; Kim, P. *Nature* **2009,** 462, (7270), 196-199.

(8) Laughlin, R. B. *Phys. Rev. Lett.* **1983,** 50, (18), 1395-1398.

(9) Jain, J. K. *Phys. Rev. Lett.* **1989,** 63, (2), 199-202.

(10) Willett, R.; Eisenstein, J. P.; Störmer, H. L.; Tsui, D. C.; Gossard, A. C.; English, J. H. *Phys. Rev. Lett.* **1987,** 59, (15), 1776-1779.

(11) Eisenstein, J. P.; Boebinger, G. S.; Pfeiffer, L. N.; West, K. W.; He, S. *Phys. Rev. Lett.* **1992,** 68, (9), 1383-1386.

(12) Suen, Y. W.; Engel, L. W.; Santos, M. B.; Shayegan, M.; Tsui, D. C. *Phys. Rev. Lett.* **1992,** 68, (9), 1379-1382.

(13) Moore, G.; Read, N. *Nucl. Phys. B* **1991,** 360, (2–3), 362-396.

(14) Ivanov, D. A. *Phys. Rev. Lett.* **2001,** 86, (2), 268-271.

(15) Peterson, M. R.; Papić, Z.; Das Sarma, S. *Phys. Rev. B* **2010,** 82, (23), 235312.

(16) Dolev, M.; Heiblum, M.; Umansky, V.; Stern, A.; Mahalu, D. *Nature* **2008,** 452, (7189), 829-834.

(17) Radu, I. P.; Miller, J. B.; Marcus, C. M.; Kastner, M. A.; Pfeiffer, L. N.; West, K. W. *Science* **2008,** 320, (5878), 899-902.

(18) Willett, R. L.; Pfeiffer, L. N.; West, K. W. *Proceedings of the National Academy of Sciences* **2009,** 106, (22), 8853-8858.

(19) Bid, A.; Ofek, N.; Inoue, H.; Heiblum, M.; Kane, C. L.; Umansky, V.; Mahalu, D. *Nature* **2010,** 466, (7306), 585-590.

(20) Venkatachalam, V.; Yacoby, A.; Pfeiffer, L.; West, K. *Nature* **2011,** 469, (7329), 185-188.

(21) Tiemann, L.; Gamez, G.; Kumada, N.; Muraki, K. *Science* **2012,** 335, (6070), 828-831.

(22) Willett, R. L. *Rep. Prog. Phys.* **2013,** 76, (7), 076501.

(23) Papić, Z.; Abanin, D. A. *Phys. Rev. Lett.* **2014,** 112, (4), 046602.

(24) Castro Neto, A. H.; Guinea, F.; Peres, N. M. R.; Novoselov, K. S.; Geim, A. K. *Rev. Mod. Phys.* **2009,** 81, (1), 109-162.

(25) Goerbig, M. O. *Rev. Mod. Phys.* **2011,** 83, (4), 1193-1243.

(26) Dean, C. R.; Young, A. F.; Cadden-Zimansky, P.; Wang, L.; Ren, H.; Watanabe, K.; Taniguchi, T.; Kim, P.; Hone, J.; Shepard, K. L. *Nat. Phys.* **2011,** 7, (9), 693-696.

(27) Feldman, B. E.; Krauss, B.; Smet, J. H.; Yacoby, A. *Science* **2012,** 337, (6099), 1196-1199.

(28) Feldman, B. E.; Levin, A. J.; Krauss, B.; Abanin, D. A.; Halperin, B. I.; Smet, J. H.; Yacoby, A. *Phys. Rev. Lett.* **2013,** 111, (7), 076802.

(29) Shibata, N.; Nomura, K. *J. Phys. Soc. Jpn.* **2009,** 78, (10).

(30) Apalkov, V. M.; Chakraborty, T. *Phys. Rev. Lett.* **2010,** 105, (3), 036801.

(31) Apalkov, V. M.; Chakraborty, T. *Phys. Rev. Lett.* **2011,** 107, (18), 186803.

(32) Papić, Z.; Abanin, D. A.; Barlas, Y.; Bhatt, R. N. *Phys. Rev. B* **2011,** 84, (24), 241306.

(33) Snizhko, K.; Cheianov, V.; Simon, S. H. *Phys. Rev. B* **2012,** 85, (20).

(34) Novoselov, K. S.; McCann, E.; Morozov, S. V.; Fal'ko, V. I.; Katsnelson, M. I.; Zeitler, U.; Jiang, D.; Schedin, F.; Geim, A. K. *Nat. Phys.* **2006,** 2, (3), 177-180.

(35) McCann, E.; Fal'ko, V. I. *Phys. Rev. Lett.* **2006,** 96, (8), 086805.

(36) Barlas, Y.; Côté, R.; Nomura, K.; MacDonald, A. H. *Phys. Rev. Lett.* **2008,** 101, (9), 097601.

(37) Abanin, D. A.; Parameswaran, S. A.; Sondhi, S. L. *Phys. Rev. Lett.* **2009,** 103, (7), 076802.

(38) Feldman, B. E.; Martin, J.; Yacoby, A. *Nat. Phys.* **2009,** 5, (12), 889-893.

(39) Zhao, Y.; Cadden-Zimansky, P.; Jiang, Z.; Kim, P. *Phys. Rev. Lett.* **2010,** 104, (6), 066801.

(40) Kharitonov, M. *Phys. Rev. Lett.* **2012,** 109, (4), 046803.

(41) Lemonik, Y.; Aleiner, I.; Fal'ko, V. I. *Phys. Rev. B* **2012,** 85, (24), 245451.

(42) Maher, P.; Dean, C. R.; Young, A. F.; Taniguchi, T.; Watanabe, K.; Shepard, K. L.; Hone, J.; Kim, P. *Nat. Phys.* **2013,** 9, (3), 154-158.

(43) Bao, W.; Zhao, Z.; Zhang, H.; Liu, G.; Kratz, P.; Jing, L.; Velasco, J., Jr.; Smirnov, D.; Lau, C. N. *Phys. Rev. Lett.* **2010,** 105, (24), 246601.

(44) Kou, A.; Feldman, B. E.; Levin, A. J.; Halperin, B. I.; Watanabe, K.; Taniguchi, T.; Yacoby, A., Electron-Hole Asymmetric Integer and Fractional Quantum Hall Effect in Bilayer Graphene. In *ArXiv e-prints*, 2013; Vol. 1312, p 7033.

(45) Skachko, I.; Du, X.; Duerr, F.; Luican, A.; Abanin, D. A.; Levitov, L. S.; Andrei, E. Y. *Phil. Trans. Roy. Soc. A* **2010,** 368, (1932), 5403-5416.

(46) Ghahari, F.; Zhao, Y.; Cadden-Zimansky, P.; Bolotin, K.; Kim, P. *Phys. Rev. Lett.* **2011,** 106, (4), 046801.

(47) Ki, D. K.; Morpurgo, A. F. *Nano Lett.* **2013,** 13, (11), 5165-70.

(48) Weitz, R. T.; Allen, M. T.; Feldman, B. E.; Martin, J.; Yacoby, A. *Science* **2010,** 330, (6005), 812-816.

(49) Velasco Jr., J.; Jing, L.; Bao, W.; Lee, Y.; Kratz, P.; Aji, V.; Bockrath, M.; Lau, C. N.; Varma, C.; Stillwell, R.; Smirnov, D.; Zhang, F.; Jung, J.; MacDonald, A. H. *Nat. Nano.* **2012,** 7, (3), 156-160.

(50) Mizuno, N.; Nielsen, B.; Du, X. *Nat. Comm.* **2013,** 4, 2716.

(51) Guimarães, M. H. D.; Veligura, A.; Zomer, P. J.; Maassen, T.; Vera-Marun, I. J.; Tombros, N.; van Wees, B. J. *Nano Lett.* **2012,** 12, (7), 3512-3517.